\documentclass[10pt,prl,aps,twocolumn,showpacs,superscriptaddress]{revtex4}
\usepackage{graphicx}

\begin{document}

\title{Nature of the antiferromagnetic to valence-bond-solid quantum phase
transition \\ in a 2D XY-model with four-site interactions}

\author{Anders W. Sandvik} 
\affiliation{Department of Physics, Boston University, 
590 Commonwealth Avenue, Boston, Massachusetts 02215}

\author{Roger G. Melko} 
\affiliation{Materials Science and Technology Division, Oak Ridge National Laboratory,
Oak Ridge, Tennessee 37831}

\date{\today}

\pacs{75.10.-b, 75.10.Jm, 75.40.Mg}

\begin{abstract}
We report large-scale quantum Monte Carlo calculations at the $T=0$ antiferromagnetic
to valence-bond-solid (VBS) transition of a two-dimensional $S=1/2$ XY model with
four-spin interactions. Finite-size scaling suggests a discontinuous spin stiffness, 
but a continuous VBS order-parameter. We propose that this is a continuous VBS 
transition without critical spin fluctuations. We also argue that the system is close 
to a point, in an extended parameter space, where both the magnetic and VBS fluctuations 
are critical---possibly a deconfined quantum-critical point.
\end{abstract}

\maketitle

Senthil {\it et al.}~recently proposed a class of generic {\it deconfined 
quantum-critical points} describing phase transitions between $O(2)$ or $O(3)$ 
antiferromagnetic (AF) and four-fold degenerate valence-bond-solid (VBS) ground 
states in two dimensions (2D) \cite{sen04}. These critical points fall outside the 
standard Landau-Ginzburg-Wilson framework, where order--order transitions are 
generically first-order (except at fine-tuned multi-critical points). A continuum 
field theory of spinons interacting with a $U(1)$ gauge field was proposed in which 
the critical point is not explicitly described in terms of order parameters. 
Instead, AF or VBS order is a consequence of confinement of spinons. This 
remarkable, but so far untested, theory calls for numerical studies of lattice 
models that could potentially exhibit deconfined quantum-criticality. It was 
suggested \cite{sen04} that such a transition may already have been observed 
in a quantum Monte Carlo (QMC) study \cite{san02a} of an XY-model including 
four-site interactions. Such a model should also from a theoretical perspective 
be among the most natural candidates for the proposed physics \cite{sac02a,sen04}. 

The J-K model studied in \cite{san02a} is defined by
\begin{equation}
H = J\sum\limits_{\langle ij\rangle} B_{ij}
    -K\sum\limits_{\langle ijkl\rangle} P_{ijkl},
\label{ham}
\end{equation}
where $B_{ij}$ and $P_{ijkl}$ are, respectively, operators acting on
nearest-neighbor sites and four sites on the corners of a plaquette
on a 2D square lattice;
\begin{eqnarray}
B_{ij} & = & S^+_iS^-_j + S^-_iS^+_j = 2(S^x_iS^x_j + S^y_iS^y_j), 
\label{bond} \\
P_{ijkl} & = & S^+_iS^-_jS^+_kS^-_l + S^-_iS^+_jS^-_kS^+_l.
\label{plaquette}
\end{eqnarray}
This $S=1/2$ model is equivalent to a half-filled lattice of hard-core bosons.
It was found to undergo a continuous AF-VBS transition at $K/J \approx 7.9$, 
with no intervening coexistence region. However, recent simulations 
of related bosonic current models have instead shown weakly first-order transitions 
\cite{kuk04}. This motivates us to revisit the J-K model and analyze the
transition in greater detail.

Using a stochastic series expansion (SSE) QMC algorithm  \cite{syl02a,mel05}, 
we have carried out extensive simulations of the J-K model in the vicinity of 
the AF-VBS transition. We have performed ground-state ($T \to 0$ converged) finite-size 
scaling for $L \times L$ lattices with $L$ up to $\approx 100$. Although the spin 
stiffness shows signs of a discontinuity, we argue that the transition is not 
necessarily first-order. We propose a scenario in which VBS order can emerge continuously 
in a state where AF order is not suppressed by diverging VBS fluctuations. The spin 
correlations change continuously from long-ranged to exponentially decaying at the 
transition, while the stiffness is discontinuous because the VBS formation is associated 
with the opening of a spin gap. While our results show that the theory of 
deconfined quantum-criticality does not apply, we argue that such a transition 
may be located nearby in an extended parameter space.

\begin{figure}
\includegraphics[width=6.5cm, clip]{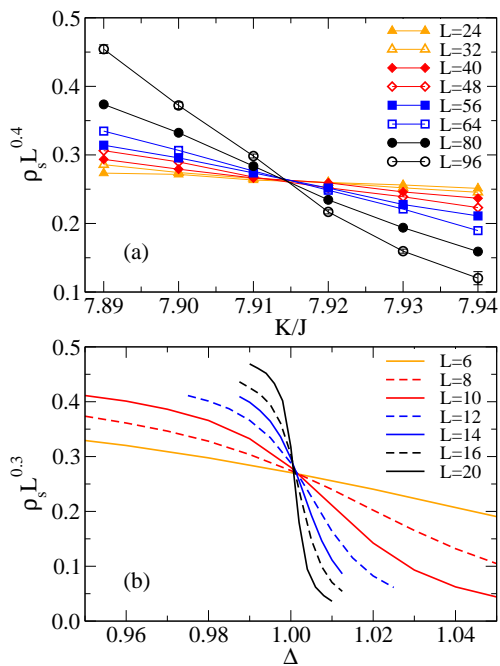}
\caption{(Color online) (a) Dependence of the size-scaled stiffness on the coupling 
$K/J$ of the J-K model, using $z=0.4$. (b) The size-scaled stiffness of the anisotropic 
Heisenberg model versus the Ising anisotropy $\Delta$, using $z=0.3$.}
\label{fig1}
\end{figure}

The spin stiffness $\rho_s$ is calculated with the the SSE method in the standard 
way \cite{mel05,pol87} in terms of winding number fluctuations. At a quantum phase 
transition with dynamic exponent $z$, it should scale with size as \cite{fis89}
\begin{equation}
\rho_s \sim L^{2-D-z} = L^{-z}.
\label{rhoscaling}
\end{equation}
We also compute the squared VBS order parameter $\langle m_P^2\rangle$ and the 
associated susceptibility $\chi_P$;
\begin{eqnarray}
\langle m_P^2\rangle & = & \frac{1}{N^2}
\sum_{a,b} \langle P_aP_b \rangle (-1)^{x_a+x_b}, \\
\chi_P & = & \frac{1}{N^2}\sum_{a,b}  \int_0^\beta d\tau
\langle P_a(\tau) P_b(0) \rangle (-1)^{x_a+x_b}.
\end{eqnarray}
Here $P_a \equiv P_{ijkl}$ and the sums are over all plaquettes. The expected 
quantum-critical  scaling is \cite{fis89}
\begin{eqnarray}
\langle m^2_P\rangle & \sim  & L^{-(z+\eta)}, \label{mscale} \\
\chi_P & \sim  & L^{-\eta}, \label{xscale}
\end{eqnarray}
where $\eta$ is the correlation function exponent. 
\goodbreak
Analyzing the stiffness according to Eq.~(\ref{rhoscaling}), we obtain $z\approx 0.4$. 
In Fig.~\ref{fig1}(a) we show how $L^{0.4}\rho_s$ versus $K/J$ graphed for different 
lattice sizes produces a crossing point at $K/J \approx 7.915$ (with $z=1$ the crossing 
points move significantly with $L$). As we will show below, this unusual value of the dynamic 
exponent is not consistent with the $T>0$ behavior of the spin susceptibility. The 
discrepancy could be interpreted as a quantum-critical point violating hyperscaling,
but a more plausible scenario is that the observed scaling is an 
artifact of the limited range of system sizes available. We therefore explore 
an alternative scenario. 

Consider the anisotropic 2D Heisenberg model;
\begin{equation}
H_{\rm Heisenberg} = J \sum_{\langle ij\rangle}
(S^x_iS^x_j + S^y_iS^y_j + \Delta S^z_iS^z_j ).
\end{equation}
Its $xy$ spin stiffness is non-zero at $T=0$ for $\Delta \le 1$ and vanishes in the 
thermodynamic limit for $\Delta > 1$. This is not due to a phase transition, but a 
consequence of the order parameter flipping from the $xy$-plane to the $z$-axis. 
Exactly at the isotropic point the stiffness should approach a constant value;
$2/3$ of the stiffness of the symmetry-broken state (reflecting rotational averaging). 
However, numerical results do not show a point at which $\rho_s$ 
becomes obviously size independent. Instead, as shown in Fig.~\ref{fig1}(b), we find 
curve crossings for $L^{0.3}\rho_s$ very close to $\Delta=1$, reminiscent of the results 
for the J-K model in Fig.~\ref{fig1}(a). This shows that a discontinuity in the 
thermodynamic limit can easily be mistaken for a continuous transition with an 
anomalously small $z$---going to very large lattices we would eventually find $z=0$.
We therefore believe that the stiffness of the J-K model is also discontinuous. This 
does not have to imply a point with enhanced symmetry. A similar behavior can 
be expected also if $\rho_s$ is discontinuous for other reason (except in the case 
of a real level crossing).

\begin{figure}
\includegraphics[width=6.5cm, clip]{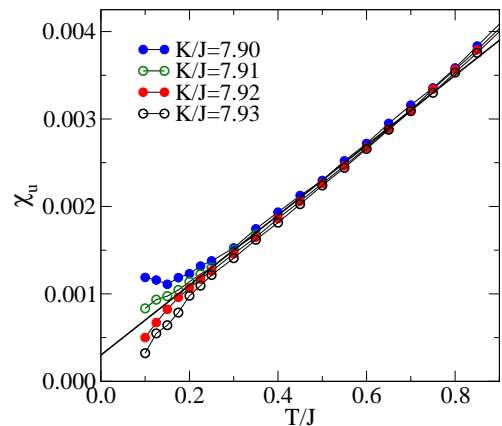}
\caption{(Color online)
Temperature dependence of the spin susceptibility for values of $K/J$ in the vicinity 
of the $T=0$ AF-VBS transition. Error bars are not shown but are at most
the same size as the symbols.}
\label{fig2}
\end{figure}

The dynamic exponent can also be extracted from the 
temperature dependence of the uniform susceptibility; $\chi_u = {J}T^{-1}\langle 
( \sum_i S^z_i )^2 \rangle/N$. For a 2D quantum-critical system it should scale as 
\citep{chu94a}
\begin{equation}
\chi_u = a + bT^{2/z-1},
\label{qcsusc}
\end{equation}
where $b$ is a constant related to the spin-wave velocity and $a=0$ at the 
critical coupling. Away from the critical coupling $a\not=0$ and there is 
low-$T$ cross-over to a different form. The earlier results \cite{san02a} 
were consistent with a $z=1$ quantum-critical point; $\chi_u \propto T$ 
at $K_c/J \approx 7.91$. We now have higher precision at lower temperatures. 
Results for $L=256$ (infinite-size converged) are shown in 
Fig.~\ref{fig2}. We observe a 
linear behavior over an extended range of temperatures, but a line fitted to the 
data has a small but clearly non-zero intercept, suggesting AF order up to 
$K_c/J = 7.915 \pm 0.005$ [note the consistency with the stiffness crossing 
point in Fig.~\ref{fig1}(a)]. Thus the AF order-parameter is not critical at the 
transition, in accord with a stiffness jump. However, the linearity of $\chi_u (T)$ 
suggests that {\it the system is close to a quantum-critical point with $z=1$ in an 
extended parameter space}. The deviations from linearity seen in Fig.~\ref{fig2} 
at low temperature are consistent with gapped spin fluctuations (exponentially
decaying $\chi_u$) for $K > K_c$ and an expected \cite{has93} cross-over 
to $T$-independence for $K \le K_c$.

\begin{figure}
\includegraphics[width=6.5cm, clip]{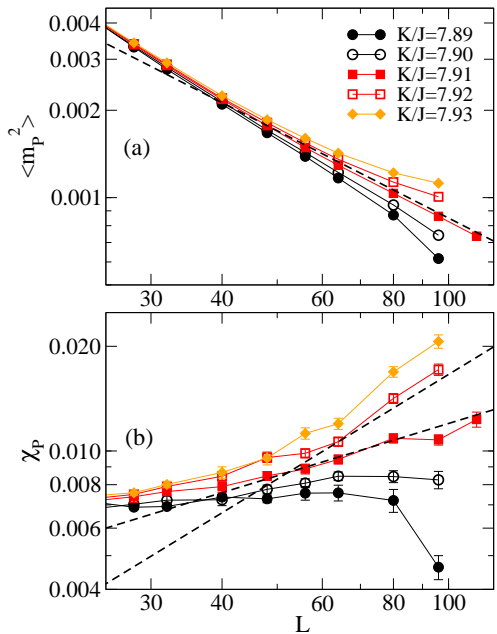}
\caption{(Color online)
Finite-size scaling of the squared VBS order parameter (a), and susceptibility 
(b). In (a) the line has slope $-1$, and the ones in (b) have slopes
$1/2$ and $1$.}
\label{fig3}
\end{figure}

Considering the discontinuities we found above, one might conclude that the
transition should be first-order. We do not find any signs of this in the
VBS correlations, however, and will therefore  consider the possibility of 
critical VBS fluctuations. Fig.~\ref{fig3} shows log-log plots of 
$\langle m_P^2\rangle$ and $\chi_P$ versus $L$ in the vicinity of the transition. 
The behaviors are typical of quantum-critical scaling. From the slope of
$\langle m_P^2\rangle$ we get $z+\eta \approx 0.98 \pm 0.04$ (taking into account 
our estimated accuracy of $K_c$) using (\ref{mscale}). The uncertainty in 
$\eta$ extracted from (\ref{xscale}) is much higher, however, due to the 
significant changes in $\chi_P$ close to $K_c$. We can only give a rough estimate, 
$\eta \in -(0.5,1.0)$, which gives $z \in (1.5,2.0)$. Since $z$ is normally 
integer, this would suggest $z=2$, but clearly further work is needed to confirm 
this. A negative $\eta$ is unusual, but the combination $z + \eta \approx 1$, 
corresponding to $\sim 1/r$ VBS correlations, is not unusual. We note that unusually 
large lattices, $L\approx 40$, are required before the (likely) asymptotic behavior 
of $\langle m_P^2\rangle$ and $\chi_P$ commences. This could be explained as a 
cross-over due to another critical point, located nearby in an extended parameter 
space, which we have already argued for above. For a first-order transition, one might 
expect such a cross-over for $K < K_c$ to be followed by a rather sharp change 
to an $1/L^2$ behavior for both $\langle m_P^2\rangle$ and $\chi_P$ 
(reflecting short-range correlations), because the cross-over should then
correspond the size of a typical VBS droplet. Instead, $\langle m_P^2\rangle$ 
for both $K < K_c$ and $K > K_c$ approaches an $\approx 1/L$ scaling, and the 
drop in $\chi_P$ occurs only for much larger $L$, well inside the AF phase. 
Although we can still not completely exclude a first-order transition, the 
intriguing possibility of critical VBS fluctuations but discontinuities in 
the spin sector at the same point deserves further analysis

\begin{figure}
\includegraphics[width=5.25cm, clip]{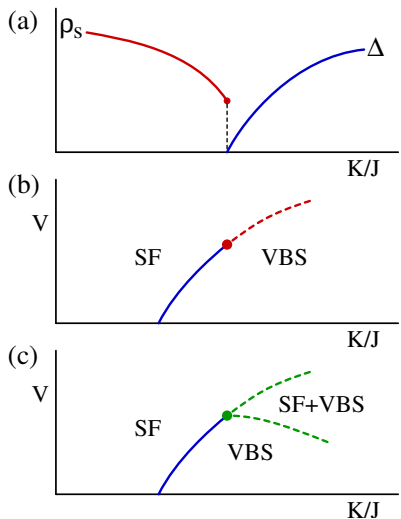}
\caption{(Color online)
(a) Conjectured behavior of the spin stiffness and the spin gap at the AF-VBS transition. 
(b),(c) Possible phase diagrams in an extended parameter space where a point with 
continuous $\rho_s$ can be reached. On the solid (blue) curves, the transition is of 
the type shown in (a), with a stiffness jump. The dashed (red) curve in (b) could 
be a line of deconfined quantum-critical points. In (c) there is an 
intervening coexistence phase.}
\label{fig4}
\end{figure}

A discontinuous $\rho_s$ can be accounted for if the VBS would be associated with 
a continuously opening spin gap, as illustrated in Fig.~\ref{fig4}(a)---the stiffness
must vanish once an infinitesimal gap opens. In this scenario the spin correlations 
would still change continuously, in a non-critical manner, from long-ranged to 
exponentially decaying. To argue for the possibility of such an exotic transition, 
we first discuss the conditions for symmetry breaking in the AF and VBS phases in terms 
of quantum levels for finite system size $N$. We then assume a continuous onset 
of VBS order and investigate the consequences of this on the relevant 
quantum states.

AF order is associated with a ``tower" of quantum-rotor states (global spin-rotation 
excitations) that become degenerate with the ground state as $1/N$ when $N \to \infty$. 
In a spin-isotropic system the ground state has spin $S=0$ and the rotor states 
$S=1,2,...$ \cite{has93}. The XY model does not conserve total spin, but we will use 
the notation $S=0,1,...$ also for its rotor states. For simplicity we restrict the 
discussion to the $S^z=0$ sector. The ground state has momentum $k=0$ and the $k>0$ 
spin wave excitations all have their own rotor towers. 

A columnar VBS state is a quadruplet corresponding to the $Z_4$ symmetry
of a columnar dimer pattern. We refer to the lattice-symmetry related 
quantum numbers with a single label, $p=0,1,2,3$, and call the states 
with $p=1,2,3$ VBS states. Like the ground state ($p=0$), the VBS states have 
$S=0$, and, due to the discrete VBS order parameter, the level spacing 
within the quadruplet vanishes exponentially with $N$. 

We now explore an AF-VBS transition involving only the states discussed above, 
i.e., those states are assumed to
stay at low energy and no other states descend from higher energies. We will follow 
the evolution of the levels $|S\rangle_p$ ($S=0,1,\ldots$; $p=0,1,2,3$) as we vary 
$K$. Although the character of the states changes as they evolve, we continue to call them 
rotor and VBS states also in their ``wrong" phase. We assume that the ground state 
$|0\rangle_0$ evolves continuously and is not crossed by any other level. 

For $\delta = K_c - K >0$ the VBS correlation length grows as $\delta^{-\nu}$, 
corresponding to a gap $\Delta \sim \delta^{z\nu}$ between $|0\rangle_0$ and $|0\rangle_1$. 
Thus, for all $K < K_c$ there is some $N$ above which there are rotor 
states $|1,2,\ldots\rangle_0$ below the first VBS state $|0\rangle_1$. 
As $N \to \infty$ a tower builds up and AF order can form. There is no apparent 
reason why the AF order has to vanish as $K \to K_c^-$. Accounting for a 
generic critical point at which both the AF and VBS orders vanish is in
fact highly non-trivial---accomplished in the theory of deconfined 
quantum-criticality \cite{sen04}. For the J-K model, our QMC results suggest
instead that the spin stiffness remains non-zero as $K \to K_c^-$, and hence 
that AF order is robust in the presence of diverging VBS correlations.

For $K > K_c$ the VBS states $|0\rangle_{123}$ approach the ground state
exponentially as $N \to \infty$. There is thus an $N$ above which the 
VBS quadruplet falls below the lowest rotor state $|1\rangle_0$. The rotor
states $|1,2,\ldots\rangle_0$ should also develop VBS correlations as $K \to K_c^-$, 
breaking the $Z_4$ symmetry for $K > K_c$. Therefore, the 
rotor states should be approached by VBS-like states $|1,2,\ldots\rangle_{123}$, 
which become degenerate at $K_c$. The tower of rotor states, along with the 
states joining them in quadruplets, thus evolve into a series of VBS-like 
states. This will also apply to the $k\not=0$ towers. What we propose 
is that these are the elementary excitations of the VBS, which hence cannot 
be regarded as ``pure VBS" excitations but involve spin as well. Clearly there 
will also be $S=0$ excitations of the VBS---these are the pure VBS 
excitations, which in our scenario would be at higher energy and irrelevant
at the transition [in contrast to deconfined quantum-criticality,
where there are gapless $S=0$ excitations due to an emergent $U(1)$ symmetry].
The key aspect of our scenario is that all the rotor states change qualitatively 
(but continuously) in a manner similar to the ground state, turning into gapped 
VBS/spin excitations. No other low-energy states emerge. With no tower of 
degenerate rotor states as $N \to \infty$ there can be no long-range AF order
and hence the stiffness jumps  to zero when the gap opens. 

What is the nature of such spinful gapped VBS excitations and, specifically,
why should they be gapped? Regarding a VBS dimer as two spins $i,j$ in a 
singlet $(\uparrow_i \downarrow_j - \downarrow_i \uparrow_j)$, the state 
$|1\rangle_{0}$ can be obtained by exciting a dimer into a state
$(\uparrow_i \downarrow_j + \downarrow_i \uparrow_j)$. This pair state can 
separate spatially into two spin-$\frac{1}{2}$ degrees of freedom---spinons---between 
which a string of out-of-phase dimers forms \cite{sachdevrmp}. The string provides 
a linear confining potential for the spinons and hence there should be a
discrete set of gapped excitations. This singlet-triplet picture is strictly based on 
$SU(2)$ symmetry, but even in the XY case, where an excited dimer is not a triplet
state, there should be analogous confining string states. At the critical point the spinons 
will become deconfined, and hence our scenario shares some features with deconfined 
quantum-criticality. There are major differences, however; the absence of emergent $U(1)$
symmetry and AF correlations that are not critical but $\sim e^{-a\Delta r}$, where 
$\Delta$ is the gap and a is a constant. The AF correlation length is thus divergent 
as $K \to K_c$ from above, but at $K_c$ the correlations are long-ranged, not power-law. 

We have not explained why the VBS forms, but we have argued that if it does form 
continuously a simultaneous AF transition of the type we have found numerically 
is not an unlikely consequence. We also cannot predict the universality class, because 
our mechanism does not require a particular scaling of the gap at the critical 
point (related to the dynamic exponent). 

We have argued for the existence, in an extended parameter space, of an AF-VBS
transition in which also the spin stiffness vanishes continuously. In Fig.~\ref{fig4}(b) 
we consider the evolution of the transition point when a suitable coupling $V$ is added to 
the J-K model. A line of critical points---which would presumably be in the deconfined 
class---could extend from the point at which the stiffness first becomes continuous. Another 
possibility is that the transition is first-order beyond this point. Fig.~\ref{fig4}(b) shows 
a scenario where there is coexisting AF and VBS order, which requires a different mechanism 
than the one we have discussed.

In principle we cannot rule out a weakly first-order AF-VBS transition \cite{kuk04,spa05}
in the J-K model, but our results do not require this. Our simulations, in combination 
with general arguments of symmetry-breaking and adiabatic evolution of quantum states, 
instead point to a new potential route to quantum phase transitions beyond the 
Ginzburg-Landau-Wilson framework. To confirm this scenario, it would clearly be 
important to identify a field theory exhibiting this type of transition.

We thank I. Affleck, L. Balents, M. P. A. Fisher, E. Fradkin, O. Motrunich, N. Prokof'ev, 
S. Sachdev, D. Scalapino, T. Senthil, B. Svistunov, and A. Vishwanath for stimulating 
discussions. AWS is supported by NSF Grant No.~DMR05-13930. We also acknowledge 
support by the NSF under Grant No.~PHY99-07949 at the KITP in Santa Barbara.

\null


\vskip-10mm
\begin{thebibliography}{00}

\bibitem{sen04}
T. Senthil, A. Vishwanath, L. Balents, S. Sachdev, and M. P. A. Fisher,
Science {\bf 303}, 1490 (2004).

\bibitem{san02a}
A. W. Sandvik, S. Daul, R. R. P. Singh, and D. J. Scalapino,
Phys. Rev. Lett. {\bf 89}, 247201 (2002).

\bibitem{sac02a}
S. Sachdev and K. Park, Annals of Physics (N.Y.) {\bf 298}, 58 (2002).

\bibitem{kuk04}
A. Kuklov, N. Prokof'ev, and B. Svistunov, Phys. Rev. Lett. {\bf 93},
230402 (2004); cond-mat/0501052.

\bibitem{syl02a} 
A. W. Sandvik, Phys. Rev. B {\bf 59}, R14157 (1999);
O. F. Sylju{\aa}sen and A. W. Sandvik, Phys. Rev. E {\bf 66}, 046701 (2002).

\bibitem{mel05} 
R. G. Melko and A. W. Sandvik, Phys. Rev. E {\bf 72}, 026702 (2005).

\bibitem{pol87}
E. L. Pollock and D. M. Ceperley, Phys. Rev. B {\bf 36}, 8343 (1987).

\bibitem{fis89}
M. P. A. Fisher, P. B. Weichman, G. Grinstein, and D. S. Fisher,
Phys. Rev. B {\bf 40}, 546 (1989). 

\bibitem{chu94a}
A. V. Chubukov, S. Sachdev and J. Ye, 
Phys. Rev. B {\bf 49}, 11919 (1994).

\bibitem{has93}
P. Hasenfratz and F. Niedermayer, Z. Phys. B {\bf 92}, 91 (1993).

\bibitem{spa05}
L. Spanu, F. Becca, and S. Sorella, cond-mat/0512272.

\bibitem{sachdevrmp}
S. Sachdev, Rev. Mod. Phys. {\bf 75}, 913 (2003).


\end{thebibliography}
\end{document}